\documentclass[prd,aps,showpacs,preprintnumbers,amssymb]{revtex4}

\usepackage{color}
\usepackage{epsf}

\begin{document}
\title{%
\hfill{\normalsize\vbox{%
\hbox{}
 }}\\
{CP violating phases and a solution to the strong CP problem }}

\author{Renata Jora
$^{\it \bf a}$~\footnote[2]{Email:
 rjora@theory.nipne.ro}}

\affiliation{$^{\bf \it a}$ National Institute of Physics and Nuclear Engineering PO Box MG-6, Bucharest-Magurele, Romania}

\date{\today}

\begin{abstract}
We present a solution to the strong CP problem based on the identification of the theta angle with twice the CP violating phase present in the CKM quark matrix. This solution washes out all the unwanted issues stemming form the strong CP phase and is strongly justified by general theoretical arguments based on the partition function associated to the $\theta$ vacuum. Phenomenological consequences on the quark mass matrices are discussed for this case.

\end{abstract}
\maketitle

\section{Introduction}

In general particle physics models are based on the invariance of the Lagrangian under symmetries which may be continuous  or discrete, global or gauged. CP symmetries is a fundamental discrete symmetry that combines parity with charge conjugation.  Early experimental data \cite{JHC} on neutral Kaon decays indicate a CP violation in this process. Later experiments on $K$ \cite{2}-\cite{4}, $B^0$ \cite{5}, \cite{6}, $B^+$ \cite{7}-\cite{9} and $B_s^0$ \cite{10} decays confirmed  this result. The nonconservation of the CP symmetry in the quark sector of the standard model  was encapsulated in the $CP$ phase  $\delta_{KM}\approx1.2$ rad \cite{PDG} of the CKM matrix \cite{CKM}.

One cannot discuss CP violation without introducing the strong CP problem. This refers specifically to QCD and is connected with  the divergence of axial current:
\begin{eqnarray}
\partial_{\mu}j^{\mu 5}=-\frac{g^2N_f}{32\pi^2}\epsilon^{\alpha\beta\mu\nu}F^a_{\alpha\beta}F^a_{\mu\nu},
\label{divaxcer84773}
\end{eqnarray}
where $N_f$ is the number of flavors and $F^a_{\mu\nu}$ is the color group gauge tensor.

The nonperturbative vacuum structure of QCD contains instanton solutions which are introduced in the Lagrangian through the term:
\begin{eqnarray}
{\cal L}_{\theta}=-\frac{\theta g^2}{32\pi^2}\epsilon^{\alpha\beta\mu\nu}F^a_{\alpha\beta}F^a_{\mu\nu}.
\label{theta537788}
\end{eqnarray}
Here $\theta$ is the vacuum angle and represents a superposition of instanton solutions with winding numbers $n$.

The strong CP problem stems mainly from the fact that $\theta$ is a strong interaction parameter but experimental data on neutron electric dipole momentum suggests that $\theta\leq10^{-9}$.  Finally the transformation from the quark gauge eigenstates basis to the mass eigenstates one  may introduce a correction to the theta term through the axial anomaly \cite{Peccei} leading to:
\begin{eqnarray}
\bar{\theta} =\theta+ {\rm arg} \det M,
\label{res62663}
\end{eqnarray}
where $M$ is the quark mass matrix.

There are various solutions \cite{Peccei} for solving the strong CP problem on dynamical grounds \cite{dynamical}, based on spontaneous CP violation \cite{SCP} or on the presence of an additional chiral symmetry \cite{axion} with or without the associated axion. Although some of the solutions are favored none of them has gathered the general acceptance.

In the present work we will propose a particular solution to the strong CP problem which amounts to the identification of the CP strong violating phase $\theta$ with twice the CP weak phase $\delta_{KM}$ (Note that a related idea but with major differences was discussed in \cite{same}). In section II we present the main theoretical ideas. Section III shows how these ideas are implemented exactly in the quark sector of the standard model. Section IV contains numerical results and the predictions of our approach. In section V we discuss the strong theoretical arguments in favor of our solution and the conclusions.

\section{Axial anomaly revisited}

We start from the QCD Lagrangian with $N_f$ flavors of fermions in the fundamental representation:
\begin{eqnarray}
{\cal L}=-\frac{1}{4}F^{a\mu\nu}F^a_{\mu\nu}+\sum_f\bar{\Psi}_fi\gamma^{\mu}D_{\mu}\Psi_f-\bar{\Psi}(C+iD\gamma^5)\Psi-\frac{\theta g^2}{32\pi^2}\epsilon^{\mu\nu\rho\sigma}F^a_{\mu\nu}F^a_{\rho\sigma},
\label{lagr35544}
\end{eqnarray}
where $D_{\mu}=\partial_{\mu}-igt^aA^a_{\mu}$ is the covariant derivative and $C$ and $D$ are hermitian matrices in the flavor space. Note that we considered the gauge eigenstate basis for further convenience. Then to the  axial transformation,
\begin{eqnarray}
&&\Psi_f\rightarrow \Psi_f+i\alpha\gamma^5\Psi_f
\nonumber\\
&&\bar{\Psi}_f\rightarrow\bar{\Psi}_f+i\alpha\bar{\Psi}_f\gamma^5,
\label{axial665774}
\end{eqnarray}
one associates the current $j_{\mu}^5$ with the anomalous conservation:
\begin{eqnarray}
\partial^{\mu}j_{\mu}^5=i\bar{\Psi}[\gamma^{5}(C+iD\gamma^5)]\Psi-\frac{g^2}{32\pi^2}\epsilon^{\mu\nu\rho\sigma}F^{a}_{\mu\nu}F^{a}_{\rho\sigma}.
\label{axanom566477}
\end{eqnarray}

In the following we shall generalize the axial transformation and make it global instead of infinitesimal. Consider the transformation:
\begin{eqnarray}
&&\Psi_{f}\rightarrow \exp[i\alpha A\gamma^5]_{fg}\Psi_g
\nonumber\\
&&\bar{\Psi}_f\rightarrow \bar{\Psi}_g\exp[i\alpha\gamma^5A]_{gf},
\label{genax535546}
\end{eqnarray}
where $A$ is hermitian matrix, $\alpha$ is an arbitrary parameter and the indices $f$ and $g$ run in the flavor space. First we will write the Lagrangian in the new variables and then the full partition function. We start with the Lagrangian (for simplicity we use the same notation for the fields):
\begin{eqnarray}
&&{\cal L}'=\bar{\Psi}\exp[i\alpha\gamma^5A]i\gamma^{\mu}\partial_{\mu}[\exp[i\alpha\gamma^5A]\Psi]+
\nonumber\\
&&g\bar{\Psi}\exp[i\alpha\gamma^5A]\gamma^{\mu}t^aA^a_{\mu}\exp[i\alpha\gamma^5A]\Psi-
\nonumber\\
&&\bar{\Psi}\exp[i\alpha\gamma^5A](C+iD\gamma^5)\exp[i\alpha\gamma^5A]\Psi=
\nonumber\\
&&\bar{\Psi}i\gamma^{\mu}iD_{\mu}\Psi+i\alpha\bar{\Psi}\gamma^5\gamma^{\mu}AiD_{\mu}\Psi+i\alpha\bar{\Psi}\gamma^{\mu}\gamma^5AiD_{\mu}\Psi+
\nonumber\\
&&-2\frac{\alpha^2}{2}\bar{\Psi}\gamma^{\mu}AiD_{\mu}\Psi-\alpha^2\bar{\Psi}\gamma^5\gamma^{\mu}\gamma^5AiD_{\mu}\Psi+....+
\nonumber\\
&&\bar{\Psi}\exp[i\alpha\gamma^5A]i\gamma^{\mu}\partial_{\mu}(\exp[i\alpha\gamma^5A])\Psi+\bar{\Psi}\exp[i\alpha\gamma^5A](C+iD\gamma^5)\exp[i\alpha\gamma^5A]\Psi=
\nonumber\\
&&\bar{\Psi}i\gamma^{\mu}iD_{\mu}\Psi+\bar{\Psi}\exp[i\alpha\gamma^5A]i\gamma^{\mu}\partial_{\mu}(\exp[i\alpha\gamma^5A])\Psi+
\nonumber\\
&&\bar{\Psi}\exp[i\alpha\gamma^5A](C+iD\gamma^5)\exp[i\alpha\gamma^5A]\Psi.
\label{resax64663}
\end{eqnarray}
Thus  the unwanted contributions from the gauge kinetic terms cancel (in each order) as they should (the dots represent the higher terms in the expansion) and the final result is quite simple.

Next we need to consider the contributions coming from the change of variable in the partition function:
\begin{eqnarray}
Z=\int \prod_f d\bar{\Psi}_fd \Psi_f \exp[i\int d^4x {\cal L}].
\label{part4566}
\end{eqnarray}
We will apply the standard Fujikawa method (see \cite{Fujikawa}). We expand the field in the gauge eigenstates $\Phi_m$ such that:
\begin{eqnarray}
&&i\gamma^{\mu}D_{\mu}\Phi_m=\lambda_m\Phi_m
\nonumber\\
&&\hat{\Phi}_m(i\gamma^{\mu}D_{\mu})=\lambda_m\hat{\Phi}.
\label{res7566477}
\end{eqnarray}
Then,
\begin{eqnarray}
&&\Psi_f(x)=\sum_ma_{mf}\Phi_m(x)
\nonumber\\
&&\bar{\Psi}_f(x)=\sum_m\hat{a}_{mf}\hat{\Phi}_m(x).
\label{chv5546636}
\end{eqnarray}
Here $f$ is the flavor index.
The functional measure then takes the form:
\begin{eqnarray}
\prod_f D\Psi_f D\bar{\Psi}_f=\prod_{mf}da_{mf}d \hat{a}_{mf},
\label{funcme64665774}
\end{eqnarray}
where upon the transformation considered in Eq. (\ref{genax535546}) the variables $a_m$ will become:
\begin{eqnarray}
a_{rm}'=\sum_n\int d^4x \Phi_m^{\dagger}(\exp[i\alpha\gamma^5A])_{rs}\Phi_na_{ns},
\label{trs77566401}
\end{eqnarray}
where again $r$, $s$ are flavor indices.
Consequently,
\begin{eqnarray}
\prod_fd\Psi_f'd \bar{\Psi}_f'=J^{-2}\prod_f d\Psi_f d\bar{\Psi}_f,
\label{transf455666}
\end{eqnarray}
where $J$ is the Jacobian with the expression:
\begin{eqnarray}
&&J=\det\Bigg[\int d^4x\Phi_m^{\dagger}(\exp[i\alpha\gamma^5A])_{rs}\Phi_n(x)\Bigg]=
\nonumber\\
&&\exp\Bigg[{\rm Tr}\ln[\int d^4x \Phi_m^{\dagger}(x)[\exp[i\alpha\gamma^5A]]_{rs}\Phi_n(x)]\Bigg].
\label{jacobttrttr}
\end{eqnarray}
Here the determinant is both in flavor and eigenvector indices space.

We will expand both the logarithm and the exponential   in order $\alpha^2$ and specialize to the case where $\alpha$ is a constant. This leads to:
\begin{eqnarray}
&&{\rm Tr} \ln[\delta^m_n+\sum_{n\geq 1}\frac{(i\alpha)^n}{n!}\int d^4x \Phi_m^{\dagger}[(\gamma^5)^nA^n_{rs}]\Phi_n\approx
\nonumber\\
&&{\rm Tr}\ln[\delta^m_n+i\alpha\int d^4 x\Phi_m^{\dagger}\gamma^5A_{rs}\Phi_n-\frac{\alpha^2}{2}\int d^4x\Phi_m^{\dagger}A^2_{rs}\Phi_n]=
\nonumber\\
&&\sum_ni\alpha\int d^4 x\Phi_n^{\dagger}\gamma^5A_{rr}\Phi_n-\frac{\alpha^2}{2}\int d^4x\Phi_n^{\dagger}(A^2)_{rr}\Phi_n+
\nonumber\\
&&\frac{\alpha^2}{2}[\int d^4 x\Phi_m^{\dagger}\gamma^5(A)_{rt}\Phi_q][\int d^4 y\Phi_q^{\dagger}\gamma^5A_{ts}\Phi_m]+....
\label{somecalc775664}
\end{eqnarray}
We further use the completness relation of the eigenstates (where $i$ and $j$ are spinor indices),
\begin{eqnarray}
\sum_n\Phi^{\dagger}_{ni}(x)\Phi_{nj}(y)=\delta(x-y)\delta_{ij}
\label{spin6764566}
\end{eqnarray}
to determine the sum of the last two terms in Eq. (\ref{somecalc775664}),
\begin{eqnarray}
&&-\frac{\alpha^2}{2}\int d^4x\Phi_n^{\dagger}(A^2)_{rr}\Phi_n+\frac{\alpha^2}{2}[\int d^4 x\Phi_{mi}^{\dagger}\gamma^5_{ij}(A)_{rt}\Phi_{qj}][\int d^4 y\Phi_{qk}^{\dagger}\gamma^5_{kl}A_{ts}\Phi_{ml}]=
\nonumber\\
&&\frac{\alpha^2}{2}(\gamma^5)^2_{ii}A^2_{rr}-\frac{\alpha^2}{2}(\gamma^5)^2_{ii}A^2_{rr}=0.
\label{res6455342}
\end{eqnarray}
Consequently in order $\alpha^2$ (but it can be shown in any order),
\begin{eqnarray}
J^{-2}=\exp[\int d^4x i\alpha \frac{g^2}{32\pi^2}\epsilon^{\mu\nu\rho\sigma}F^a_{\mu\nu}F^a_{\mu\nu}{\rm Tr}A].
\label{finalres663}
\end{eqnarray}
Here we use standard calculations of the linear term in the Fujikawa method.

Thus the partition function will become upon the change of variable:
\begin{eqnarray}
Z =\prod_f \int d\bar{\Psi}_fd\Psi_f\exp[i\int d^4x{\cal L}'],
\label{partfunc6455}
\end{eqnarray}
where,
\begin{eqnarray}
{\cal L}'&=&\sum_f\bar{\Psi}_fi\gamma^{\mu}D_{\mu}\Psi_f+i\bar{\Psi}\exp[i\alpha\gamma^5A]\gamma^{\mu}\partial_{\mu}[\partial_{\mu}\exp[i\alpha\gamma^5A]]\Psi+
\nonumber\\
&&-\bar{\Psi}\exp[i\alpha\gamma^5A](C+iD\gamma^5)\exp[i\alpha\gamma^5A]\Psi+
\nonumber\\
&&\frac{g^2}{32\pi^2}\alpha \epsilon^{\mu\nu\rho\sigma}F^a_{\mu\nu}F^a_{\rho\sigma}{\rm Tr}A-\frac{g^2}{32\pi^2}\theta\epsilon^{\mu\nu\rho\sigma}F^a_{mu\nu}F^a_{\rho\sigma}.
\label{res553442}
\end{eqnarray}
We require that the global axial transformation cancel the $\theta$ term which amounts to the constraint(Note that we may choose $\alpha$ at our discretion): $2\alpha{\rm Tr}A=\theta$. Here the factor of $2$ takes into account that the transformation in Eq. (\ref{genax535546}) with the matrix $A$ in the flavor space must be applied to both up and down quarks. We can also dismiss the total derivative that appears in the second term on the right hand side of Eq. (\ref{res553442}) to obtain that the net effect of the theta term leads to a Lagrangian which is completely equivalent to the initial one with the theta term:
\begin{eqnarray}
{\cal L}=\sum_f\bar{\Psi}_fi\gamma^{\mu}D_{\mu}\Psi_f+\bar{\Psi}\exp[i\theta\gamma^5B](C+iD\gamma^5)\exp[i\theta\gamma^5B]\Psi,
\label{finallagrt56647}
\end{eqnarray}
where we redefine $B=\frac{A}{{\rm Tr}A}$.

\section{Connection between CP violation phase and the theta angle}

We start with the quark mass terms in the standard model Lagrangian:
\begin{eqnarray}
{\cal L}=(\bar{u}_{LA}M^u_{AB}u_{RB}+\bar{d}_{LA}m^d_{AB}d_{RB})+h.c.,
\label{quarkmasst6647756}
\end{eqnarray}
where $A$, $B$ denote flavor indices for the up ($u$) and down ($d$) quarks, $L$ and $R$ denote left and right handed states. Moreover $M^u$ and $M^d$ are the mass matrices for the up and down quarks. These are $3\times 3$ arbitrary complex matrices that do not have to be hermitian. In general they are diagonalized by a biunitary transformation:
\begin{eqnarray}
&&M^u_{diag}=V_uM^UW_u^{\dagger}
\nonumber\\
&&M^d_{diag}=V_dM^dW_d^{\dagger}.
\label{masmatrxc6479}
\end{eqnarray}
Here the transformation from the gauge eigenstates (unprimed fields) to the mass eigenstates (primed fields) is given by:
\begin{eqnarray}
&&u_L'=V_u u_L
\nonumber\\
&&d_L'=V_d d_L
\nonumber\\
&&u_R'=W_u u_R
\nonumber\\
&&d_R'=W_d d_R.
\label{transform664553}
\end{eqnarray}

The standard approach is to first make the change of variables in the left sector:
\begin{eqnarray}
&&u_L'=V_uu_L
\nonumber\\
&&\bar{u}_L'=\bar{u}_LV_u^{\dagger}
\nonumber\\
&&d_L"=V_uu_L
\nonumber\\
&&\bar{d}_L"=\bar{d}_LV_u^{\dagger}.
\label{chnagevar4665}
\end{eqnarray}
Then the up quark mass matrix will be diagonalized by:
\begin{eqnarray}
V_uM^uM^{u\dagger}V_u^{\dagger}=M^{u 2}_{diag},
\label{resc546363}
\end{eqnarray}
and one further needs to diagonalize the new down quark mass matrices:
\begin{eqnarray}
V_uM^dM^{d\dagger}V_u^{\dagger}.
\label{newms65774}
\end{eqnarray}
It turns out that this new matrix is diagonalized by $U=V_uV_d^{\dagger}$ where $U$ is the CKM matrix:
\begin{eqnarray}
U^{\dagger}V_uM^dM^{d\dagger}V_u^{\dagger}U=M^{d2}_{diag}.
\label{res6645}
\end{eqnarray}
 This corresponds to the transformation $d_L"=V_uV_d^{\dagger}d_L'$. In the following we will use:
\begin{eqnarray}
V_uM^dM^{d\dagger}V_u^{\dagger}=UM^{d2}_{diag}U^{\dagger},
\label{impty56657}
\end{eqnarray}
derived directly from Eq. (\ref{res6645}).

We  make the following important assumption $W_u^{\dagger}=V_u$. Now consider the axial transformation discussed in section II with $2\alpha{\rm Tr}A=\theta$:
\begin{eqnarray}
u'=\exp[i\frac{\theta}{2} B\gamma^5]u.
\label{transfrom65774}
\end{eqnarray}
This can also be written as:
\begin{eqnarray}
&&\left(
\begin{array}{c}
u_L'\\
u_R'
\end{array}
\right)=(1+i\frac{\theta}{2} B\gamma^5-\frac{\theta^2}{8}B^2+...)\left(
\begin{array}{c}
u_L\\
u_R
\end{array}
\right)=
\nonumber\\
&&
\left(
\begin{array}{c}
(1-i\frac{\theta}{2} B-\frac{\theta^2}{8}B^2+...)u_L\\
(1+i\frac{\theta}{2} B-\frac{\theta^2}{8}B^2+...)u_R
\end{array}
\right)=
\left(
\begin{array}{c}
\exp[-i\frac{\theta}{2} B]u_L\\
\exp[i\frac{\theta}{2} B]u_R
\end{array}
\right).
\label{Matrixeq663554}
\end{eqnarray}
Then with the identification $V_u=\exp[-i\frac{\theta}{2} B]$ we obtain exactly the transformation in Eq. (\ref{transform664553}). Then the axial transformation we considered in the first section is not just put by hand but it is the required change  from the gauge eigenstate basis to the mass eigenstate one.

The first consequence of our assumption is the expression for the mass matrix for the up quarks:
\begin{eqnarray}
M^u=V_u^{\dagger}M^u_{diag}V_u^{\dagger}=\exp[i\theta B]M^d_{diag}\exp[i\theta B].
\label{firstrel}
\end{eqnarray}
Next we know:
\begin{eqnarray}
U^{\dagger}V_uM^dV_u(V_u^{\dagger}W_d^{\dagger})=M^d_{diag},
\label{tworel775664}
\end{eqnarray}
so we require  $W_d=U^{\dagger}V_u^{\dagger}$ such that no new axial transformation of the fermion fields is introduced. From the definition of the CKM matrix one has $V_d=U^{\dagger}V_u$ so the final mass matrix for the down quarks can be computed as:
\begin{eqnarray}
M^d=V_d^{\dagger}M^d_{diag}W_d.
\label{secexpr664553}
\end{eqnarray}

At this point all we need is to calculate $V_u=\exp[-i\frac{\theta}{2} B]$ form Eq. (\ref{impty56657}). For that we write:
\begin{eqnarray}
M^d=C+iD.
\label{rel8856647}
\end{eqnarray}
 We first consider $\frac{\theta}{2}=\delta_{KM}$  arbitrary and expand the left hand and right hand sides of Eq. (\ref{impty56657}) in order  $\delta_{KM}^2$. We denote:
\begin{eqnarray}
M^dM^{d\dagger}=M_1+iM_2=UM^{d2}_{diag}U^{\dagger}|_{\delta_{KM}=0}=M_1.
\label{not10483}
\end{eqnarray}
In first and second orders in $\delta_{KM}$ we get:
\begin{eqnarray}
&&iS_1=-iBM_1+iM_1B=\frac{\partial UM^{d2}_{diag}U^{\dagger}}{\partial \delta}|_{\delta_{KM}=0}
\nonumber\\
&&BM_1B-\frac{1}{2}(B^2M_1-M_1B^2)=\frac{1}{2}\frac{\partial^2 UM^{d2}_{diag}U^{\dagger}}{(\partial \delta_{KM})^2}|_{\delta_{KM}=0}.
\label{relveryomori5746636}
\end{eqnarray}
The last equation in Eq. (\ref{relveryomori5746636}) can be simplified as:
\begin{eqnarray}
-\frac{1}{2}(SB_1-S_1B)=\frac{1}{2}\frac{\partial^2 UM^{d2}_{diag}U^{\dagger}}{(\partial \delta_{KM})^2}|_{\delta_{KM}=0}.
\label{res64553}
\end{eqnarray}

\section{Numerical calculations}
We consider the standard expression for the CKM matrix:
\begin{eqnarray}
U=\left(
\begin{array}{ccc}
c_{12}c_{13}&s_{12}c_{13}&s_{13}\exp[-i\delta]\\
-s_{12}c_{23}-c_{12}s_{23}s_{13}\exp[i\delta]&c_{12}c_{23}-s_{12}s_{23}s_{13}\exp[i\delta]&s_{23}c_{13}\\
s_{12}s_{23}-c_{12}c_{23}s_{13}\exp[i\delta]&-c_{12}s_{23}-s_{12}c_{23}s_{13}\exp[i\delta]&c_{23}c_{13}
\end{array}
\right),
\label{umatrix6465536}
\end{eqnarray}
with,
\begin{eqnarray}
&&s_{12}=\frac{|V_{us}|}{\sqrt{|V_{ud}|^2+|V_{us}|^2}}
\nonumber\\
&&s_{23}=s_{12}\frac{|V_{cb}|}{|V_{us}|}
\nonumber\\
&&s_{13}=|V_{ub}|.
\label{res6635252}
\end{eqnarray}
Here $|V_{us}|=0.2248$, $|V_{ud}|=0.97417$, $|V_{cb}|=40.5\times 10^{-3}$, $|V_{ub}|=-4.09\times 10^{-3}$ (\cite{PDG}).

The matrix equations (\ref{relveryomori5746636}) contain  $9$ algebraic equations. From the first matrix we determine that $B$ which is in general a hermitian matrix must be real and symmetric and thus contains $6$ unknown parameters. If we add the trace condition ${\rm Tr}B=1$ we have a system of $10$ equations with $6$ unknowns. It turns out that this system  has a very good approximate  solution which means that four equations are automatically satisfied almost exactly. This matching is a very strong argument in favor of the correctitude of our initial assumption $V_u=\exp[i\theta B]$. Finally the matrix $B$ is given by:
\begin{eqnarray}
B =\left(\begin{array}{ccc}
0.6667&3.2848\times 10^{-9}&-1.3321\times 10^{-10}\\
3.2848\times 10^{-9}&0.6665&-0.0405\\
-1.3321\times 10^{-10}&-0.0405&-0.3317
\end{array}
\right)
\label{bmatr663773}
\end{eqnarray}

From Eq. (\ref{firstrel}) we calculate the mass matrix for the up quarks as:
\begin{eqnarray}
&&M^u=
\nonumber\\
&&\left(
\begin{array}{ccc}
-0.00005+0.00219i&-4.96914\times 10^{-9}-1.78004\times  10^{-10}i&1.20997\times 10^{-10}+1.92299\times 10^{-10}i\\
6.62363-4.27139i&-0.02788+1.27275i&0.04888-0.03163i\\
150.035+67.135i&0.04888-0.03163i&-0.00245-0.00104i
\end{array}
\right).
\label{upqaumd66464}
\end{eqnarray}

Further on the down quark mass matrix is determined from Eq. (\ref{secexpr664553}).
\begin{eqnarray}
M^d=
\left(
\begin{array}{ccc}
-0.00025+0.01016i&0.00121+0.01815i&0.04204-0.04425i\\
-0.00056+0.02237i&0.00250+0.08597i&0.11778-0.12483i\\
-0.00162+0.05961i&0.11799-0.12533i&2.90708-2.99229i
\end{array}
\right).
\label{downqy5664553}
\end{eqnarray}

\section{Conclusions}

In this work we presented a theoretical idea and its phenomenological implementation for solving the strong CP problem.  This was based on cancelling the full theta angle $\bar{\theta}$  in Eq. (\ref{res62663}) and on identifying the instanton $\theta$ angle with twice the CP violation phase in the CKM matrix. We showed how this works and what are the consequences. All other dependence on $\theta$ in the Lagrangian besides the CKM matrix is thus washed out.  Moreover since $\theta$ is of order $2$ it makes  perfect sense in a theory of strong interaction.

 In the end we will show that our solution although particular makes perfect sense and it is completely justifiable in some theoretical context.

Consider that only the up quark mass matrix has a complex determinant such that ${\rm arg}\det M^u=\alpha$. Then one can make a phase transformation on the right handed up quark states such that this phase is eliminated from the mass matrix and introduced in $\bar{\theta}=\theta+\alpha$. Then the whole process of determining the CKM matrix seems independent of $\alpha$. However there are situations when this conclusion is simply not true.

 In order to show that there are instances where the strong phase may remain relevant for the CKM matrix let us consider the following hypothetical example. Assume  that there is some flavor symmetry relating the elements of $M^u$ such that $M^u=PX^u$ where $X^u$ is a real symmetric matrix and:
\begin{eqnarray}
P=
\left(
\begin{array}{ccc}
\exp[i\alpha_1(c)+i\frac{\alpha(c)}{3}]&0&0\\
0&\exp[-i\alpha_1(c)+i\frac{\alpha(c)}{3}]&0\\
0&0&\exp[i\frac{\alpha(c)}{3}]
\end{array},
\right)
\label{phase87758}
\end{eqnarray}
where $c$ is a parameter depending on the group structure  and $\alpha(c)={\rm arg}\det[M^u]$. One may eliminate the phase $\alpha(c)$ as discussed previously and determine:
\begin{eqnarray}
&&V_u=V_u'P^*
\nonumber\\
&&U_{CKM}=V_u'P^*V_d.
\label{contex6564773}
\end{eqnarray}
Here $V_u'$ is real and unitary and we may assume that also $V_d$ has the same properties. Then $\alpha_1(c)$ is simply the  CKM phase as it is the only phase present in the CKM matrix. But in the presence of the flavor symmetry both $\alpha_1(c)$ and $\alpha(c)$ depend on $c$ and it is possible to eliminate $c$ in favor of $\alpha$ which leads to  $\alpha_1(\alpha)$. Consequently the strong phase $\alpha$ or a function of it becomes a  physical phase directly related to the $CKM$ phase.

Our arguments are however more powerful than that. Next we will show  that the fact that $\alpha$ becomes a physical phase means that $\bar{\theta}=\theta+\alpha$ becomes irrelevant in the Lagrangian.

For a $\theta$ vacuum defined as:
\begin{eqnarray}
|\theta \rangle=\sum_n\exp[i\theta n]|n\rangle,
\label{res63554}
\end{eqnarray}
the associated partition function is:
\begin{eqnarray}
Z_{\theta}=\sum_n\exp[i\theta n]\int dA_n d\bar{\Psi}d\Psi \exp[iS(A_n,\bar{\Psi},\Psi)],
\label{inst5535}
\end{eqnarray}
where $A_n$ represent all gauge field solution in a class corresponding to the winding number $n$ and $\Psi$ all standard model quarks.
The transformation from the gauge eigenstate basis to the mass eigenstate    introduces in the most general case an axial transformation and through axial anomaly  a change of the theta angle to:
\begin{eqnarray}
\theta\rightarrow \theta+\alpha+\beta,
\label{res7745553}
\end{eqnarray}
where $\alpha$ is the up quarks contribution and $\beta$ is the down quarks one. The mass matrices become diagonal and the angles $\alpha$ and $\beta$ may appear only in the CKM matrix. We write:
\begin{eqnarray}
Z_{\theta,\alpha,\beta}=\sum_n\exp[in(\theta+\alpha+\beta)]f_n(\alpha,\beta),
\label{res66377288}
\end{eqnarray}
where $f_n$ is just a notation for the part of the partition function independent on $\theta$ and where the quark mass matrices are diagonal without any phases.
Then by Fourier analysis:
\begin{eqnarray}
f_n(\alpha,\beta)=\frac{1}{2\pi}\int_{-\pi}^{\pi}Z_{\theta,\alpha,\beta}\exp[-in(\theta+\alpha+\beta)]d \theta .
\label{res663552}
\end{eqnarray}
Let us consider that $\beta=0$ and $\alpha$ is a physically phase relevant for the CKM matrix. Then we can differentiate Eq. (\ref{res663552}) with respect to $\alpha$ to get (Note that the fact that $\alpha$ is a physical parameter is crucial for this step):
\begin{eqnarray}
&&\frac{\partial f_n}{\partial \alpha}=
\frac{1}{2\pi}\Bigg[\int_{-\pi}^{\pi}\frac{\partial Z_{\theta, \alpha}}{\partial \alpha}\exp[-in(\theta+\alpha)]d\theta+\int_{-\pi}^{\pi}(-in)Z_{\theta,\alpha}\exp[-in(\theta+\alpha)]\Bigg]=
\nonumber\\
&&\frac{1}{2\pi}\Bigg[\int_{-\pi}^{\pi}\frac{\partial Z_{\theta, \alpha}}{\partial \alpha}\exp[-in(\theta+\alpha)]d\theta+
\int_{-\pi}^{\pi}Z_{\theta,\alpha}\frac{\partial}{\partial \theta}\exp[-in(\theta+\alpha)]\Bigg]=
\nonumber\\
&&\frac{1}{2\pi}\Bigg[\int_{-\pi}^{\pi}\frac{\partial Z_{\theta, \alpha}}{\partial \alpha}\exp[-in(\theta+\alpha)]d\theta-\int_{-\pi}^{\pi}\frac{\partial Z_{\theta,\alpha}}{\partial \theta}\exp[-in(\theta+\alpha)]\Bigg].
\label{calcfinstr66488}
\end{eqnarray}
Here we integrated by parts in the second term of the right hand side of the equation and used:
\begin{eqnarray}
Z_{\pi,\alpha}\exp[-in(\pi+\alpha)]=Z_{-\pi,\alpha}\exp[-in(-\pi+\alpha)].
\label{res64553}
\end{eqnarray}
The partition function is independent of an axial transformation performed both in the action and the variables of integration. We then consider  a transformation on the right handed up states (by $\exp[-i\alpha]$) and down quark states (by $\exp[-i\theta]$) such that the relevant part of the Lagrangian becomes:
\begin{eqnarray}
{\cal L}_m=\bar{u}_LM^u\exp[-i\alpha]u_R+\bar{d}_LM^d\exp[-i\theta]u_R+h.c.
\label{res63552}
\end{eqnarray}
This eliminates completely the instanton term proportional to the dual tensor in the Lagrangian.
We sum over $n$ on the right and left handed sides of the Eq. (\ref{calcfinstr66488}) and use Eq. (\ref{res63552}) to obtain:
\begin{eqnarray}
&&\int d\bar{\Psi} d\Psi d A^a_{\mu}\Bigg[i\int d^4x[-\frac{g}{\sqrt{2}}\bar{u}_L\gamma^{\mu}W_{\mu}^{+}\frac{ \partial U_{CKM}}{ \partial \alpha}d_L+h.c.] \Bigg]\exp[iS]=
\nonumber\\
&&\int d \bar{\Psi} d\Psi d A^a_{\mu}\frac{1}{2\pi}\int_{-\pi}^{\pi} d\theta \Bigg[i\int d^4x[-i\bar{u}_LM^u\exp[-i\alpha]u_R+i\bar{d}_LM^d\exp[-i\theta]u_R+h.c.]\Bigg]\delta(\alpha+\theta)\exp[iS]=
\nonumber\\
&&\int d \bar{\Psi} d\Psi d A^a_{\mu}\Bigg[i\int d^4x[-i\bar{u}_LM^uu_R++i\bar{d}_LM^du_R+h.c.]\Bigg]\exp[iS]
\label{res77466}
\end{eqnarray}
Here we integrated over $\theta$ on the right hand side and perform again an axial transformation to eliminate the phases in the second line of Eq. (\ref{res77466}).
The final result  in Eq. (\ref{res77466}) has no trace of strong CP phase and relates two distinct independent operators in the Lagrangian. The only way to make sense of this result is to conclude that in this case the quantity $\bar{\theta}=\alpha+\theta$ is completely  irrelevant  and that one can safely take $\bar{\theta}=0$. This solves the strong CP problem.

If $\alpha=\beta=0$ our procedure would  not apply. However this case is highly improbable in the standard model since electroweak loops introduce with certainty $\gamma^5$ corrections in the quark masses even in the absence of the theta term.

In summary we proposed a particular approximate solution to the strong CP  problem that eliminates all major issues related to it and that is perfectly justified on more general theoretical grounds.  We also discussed the phenomenological consequences in the quark sector of the standard model. Other applications of our approach together with a more comprehensive numerical solution will be discussed elsewhere.


\begin{thebibliography}{30}
\bibitem{JHC} J. H. Christensen et al., Phys.Rev. Lett. {\bf 13}, 138 (1964).
\bibitem{2} H. Burkhardt et al. [NA31 Collab], Phys. Lett. B {\bf 26}, 169 (1988).
\bibitem{3} V. Fanti et al. [NA48 Collab.], Phys. Lett. B {\bf 465}, 335 (1999).
\bibitem{4} A. Alavi-Harati et al. [KTeV Collab.], Phys. Rev. Lett. {\bf 83}, 22 (1999).
\bibitem{5} B. Aubert et al. [Babar Collab.], Phys. Rev. Lett. {\bf 93}, 131801 (2004).
\bibitem{6} Y. Chao et al. [Belle Collab.], Phys. Rev. Lett. {\bf 93}, 191802 (2004).
\bibitem{7} A. Poluektov et al.. [Belle Collab.], Phys. Rev. D {\bf 81}, 112002 (2010).
\bibitem{8} P. del Amo Sanchez et al. [BABAR Collab.], Phys. Rev. D {\bf 82}, 072004 (2010).
\bibitem{9} R. Aaij et al. [LHCb Collab.], Phys. Lett. B {\bf 712}, 203 (2012).
\bibitem{10} R. Aaij et al. [LHCb Collab.], Phys. Rev. Lett.  {\bf 110}, 221601 (2013).
\bibitem{PDG} C. Patrignani et al.  (Particle Data Group), Chin. Phys. C {\bf 40}, 100001 (2016).
\bibitem{CKM} M. Kobayashi and T. Maskawa, Progr. Theor. Phys. {\bf 49}, 652 (1973).
\bibitem{Peccei} R. D. Peccei, Lect. Notes. Phys. {\bf 741}, 3-17 (2008).
\bibitem{dynamical} S. Khlebnikov and M. Shaposhnikov, Phys. Lett. B {\bf 203}, 121 (1988).
\bibitem{SCP} Y. B. Zeldovich, I. B. Kobzarev and L. Okun, Sov. Phys. JTEP {\bf 40}, 1 (1975).
\bibitem{axion} R. D. Peccei and H. R. Quinn, Phys. Rev. Lett. {\bf 38}, 1440 (1977); Phys. Rev. D {\bf 16}, 1791 (1977).
\bibitem{same} J. Bordes, H. M. Chan, S. T. Tsou, Int. J. Mod. Phys. A {\bf 25} 5897-5911 (2010).
\bibitem{Fujikawa} K. Fujikawa, Phys. Rev. D {\bf 29}, 285 (1984).
\end{thebibliography}
\end{document}